\documentclass[conference]{IEEEtran}
\IEEEoverridecommandlockouts

\usepackage{cite}
\usepackage{amsmath,amssymb,amsfonts}
\usepackage{algorithmic}
\usepackage{graphicx}
\usepackage{textcomp}
\usepackage{xcolor}
\usepackage[mathscr]{eucal}
\usepackage{multirow}
\usepackage{bbding}
\usepackage{booktabs}
\usepackage{hyperref} 
\hypersetup{hidelinks}
\def\BibTeX{{\rm B\kern-.05em{\sc i\kern-.025em b}\kern-.08em
    T\kern-.1667em\lower.7ex\hbox{E}\kern-.125emX}}
\begin{document}

\title{Cascading Relevance-driven Recommendation
Network for CTR Prediction in Trigger-Introduced Recommendation\\
}

\author{\IEEEauthorblockN{1\textsuperscript{st} Kaixuan Chen}
\IEEEauthorblockA{\textit{Taobao \& Tmall Group of Alibaba} \\
Hangzhou, China \\
chenkaixuan.ckx@taobao.com}
\and
\IEEEauthorblockN{2\textsuperscript{nd} Wenwen Wang}
\IEEEauthorblockA{\textit{Taobao \& Tmall Group of Alibaba} \\
Hangzhou, China \\
www234574@taobao.com}
\and
\IEEEauthorblockN{3\textsuperscript{rd} Xing Fang}
\IEEEauthorblockA{\textit{Taobao \& Tmall Group of Alibaba} \\
Hangzhou, China \\
fangxing.fx@taobao.com}
\and
\IEEEauthorblockN{4\textsuperscript{th} Yang Huang}
\IEEEauthorblockA{\textit{Taobao \& Tmall Group of Alibaba} \\
Hangzhou, China \\
21631140@zju.edu.cn}
\and
\IEEEauthorblockN{5\textsuperscript{th} Jing Wang}
\IEEEauthorblockA{\textit{Taobao \& Tmall Group of Alibaba} \\
Hangzhou, China \\
jing.wangj1@taobao.com}
}

\maketitle

\begin{abstract}
E-commerce has emerged as crucial platforms for people's daily consumption and shopping interests. There is a new recommendation scenario, Trigger-Introduced Recommendation (TIR), where users click interested product, which is defined as the trigger item, containing their instant interest, and in the undertaking page following the relevant target items. Distinguished from traditional search and recommendation scenarios, trigger contains relatively strong instant interest, which is more vague and implicit compared to search terms. Relying on large amounts of labeled data, existing methods lack the exploration of trigger relevance, which affects users' immersive experience. To alleviate this problem, we propose the Cascading Relevance-driven Recommendation Network (CRRN) to emphasize the interaction and relevance between trigger and target, comprising three essential components: 1) the Trigger-Target Interaction layer extracts interaction features of trigger and target based on personalized gating. 2) Cascading Interest Fusion module explicitly estimates users' trigger intention and fuses instant and personalized interests adaptively with cascading attention blocks. 3) Category-assisted Pairwise Loss enhances trigger relevance with the guidance of category association between trigger and target. Extensive experiment results show that CRRN outperforms recent state-of-the-art methods on both industrial and public datasets. Online A/B tests further validate the effectiveness of our method. Our code is available\footnote{\url{https://github.com/a-little-cabbage/CRRN}}.
\end{abstract}


\section{Introduction}
\begin{figure}[htpb]
\centering
\includegraphics[scale=0.33]{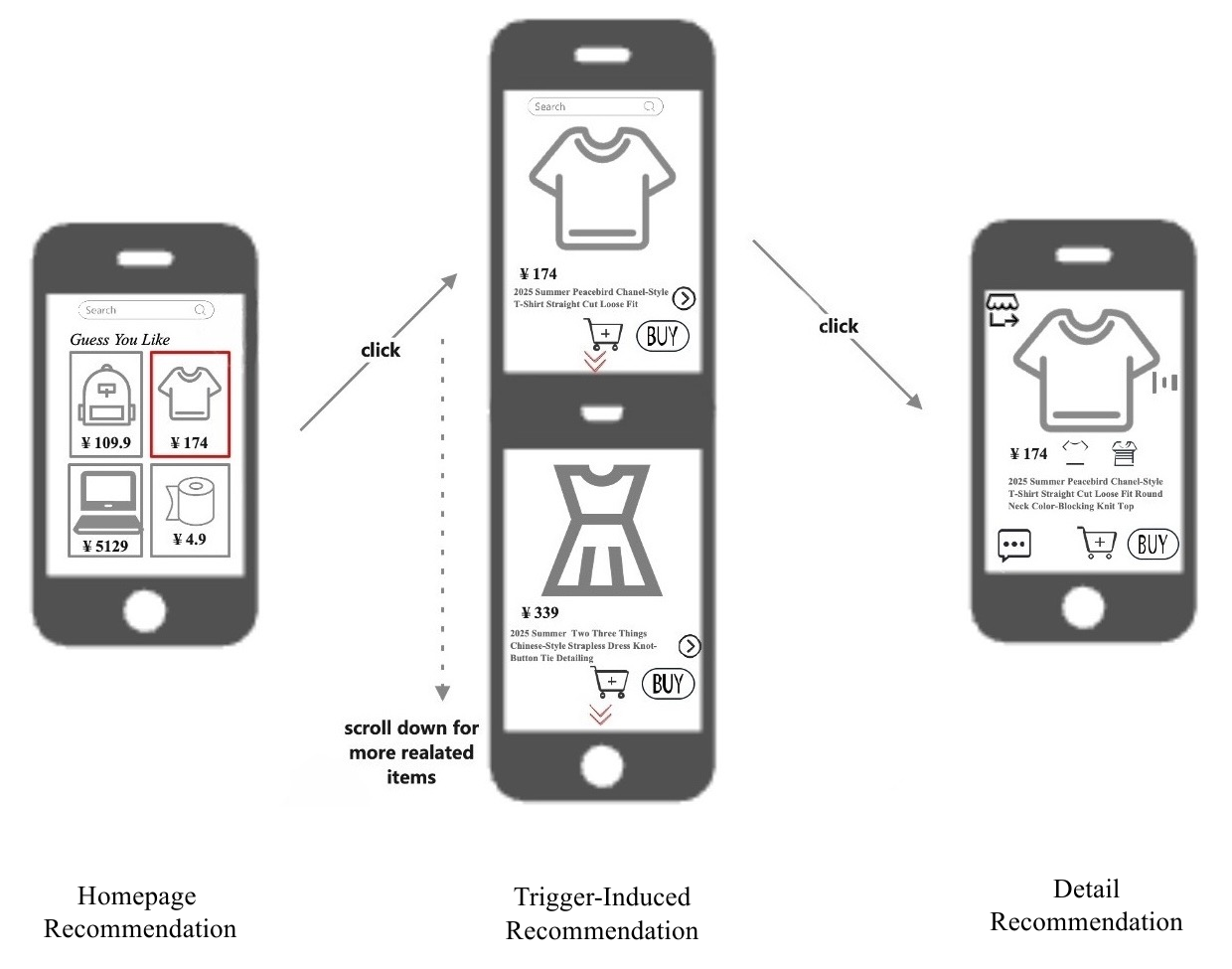}
\caption{An illustration of the TIR scenario. Left: Homepage Recommendation, middle: Trigger-Induced Recommendation, right: Detail Recommendation.}
\label{intro}
\end{figure}

Trigger-Induced Recommendation has become increasingly important in the e-commerce platforms \cite{platform1}, delivering an immersive, focused online shopping experience for users. As shown in Fig \ref{intro}, when users click on a product from the homepage, which is defined as the trigger item, it triggers a new scenario named Trigger-Induced Recommendation shown in the middle figure, displaying a series of related products. Users can scroll through this page to view additional items, and if they are interested in a specific product, they can click on it to access its detail page as shown on the right. 
Traditional recommendation methods \cite{ctr1,ctr2,ctr3,ctr4,ctr5}, such as user-induced recommendation, mainly explore users' preferences based on historical behaviors \cite{behavior1,behavior2,behavior3} by attention mechanisms \cite{attn1,attn2}. However, trigger serves as a unique entity containing strong signals of instant interest in TIR scenarios, and should be considered specially and jointly processed with personalized interests.

\begin{figure*}[htpb]
\centering
\includegraphics[scale=0.5]{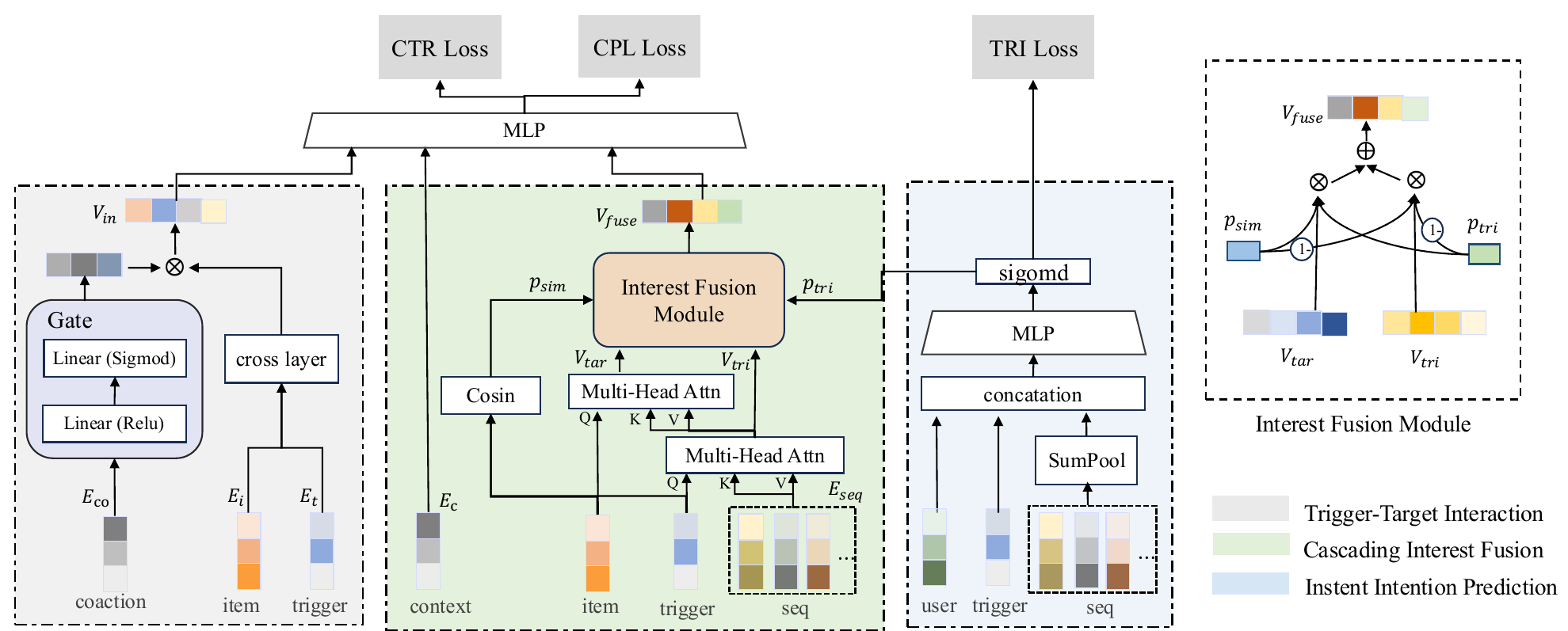}
\caption{The architecture of the proposed CRRN.}
\label{model}
\end{figure*}

Current state-of-the-art methods \cite{tri1,DIHN,DIAN, DEI2N} in industrial recommendation systems, such as DIN \cite{DIN} and SIM \cite{SIM}, fail to explicitly utilize trigger features, rendering it insufficient for delivering an immersive user experience in TIR scenarios. Despite the growing importance of TIR in industry, relevant research remains limited. Tencent's R3S \cite{r3s} captures immediate user interests through feature interactions, semantic similarity, and information gain, but neglects the critical feature of user historical behavior sequences. DIHN \cite{DIHN} proposes an interest highlighting network that predicts users' intentions and fuses features of trigger-relevant and target-relevant from users' historical behaviors. DIAN \cite{DIAN} estimates the user's intention through an intent-aware network, distinguishing between trigger-aware and trigger-free features and performing weight fusion. DEI2N \cite{DEI2N} focuses on exploiting temporal information and dynamic changes of the trigger intention during page scrolling. The above methods make great efforts to user's trigger intention while neglecting the enhancement of trigger relevance learning.  

\begin{figure}[htpb]
\centering
\includegraphics[scale=0.25]{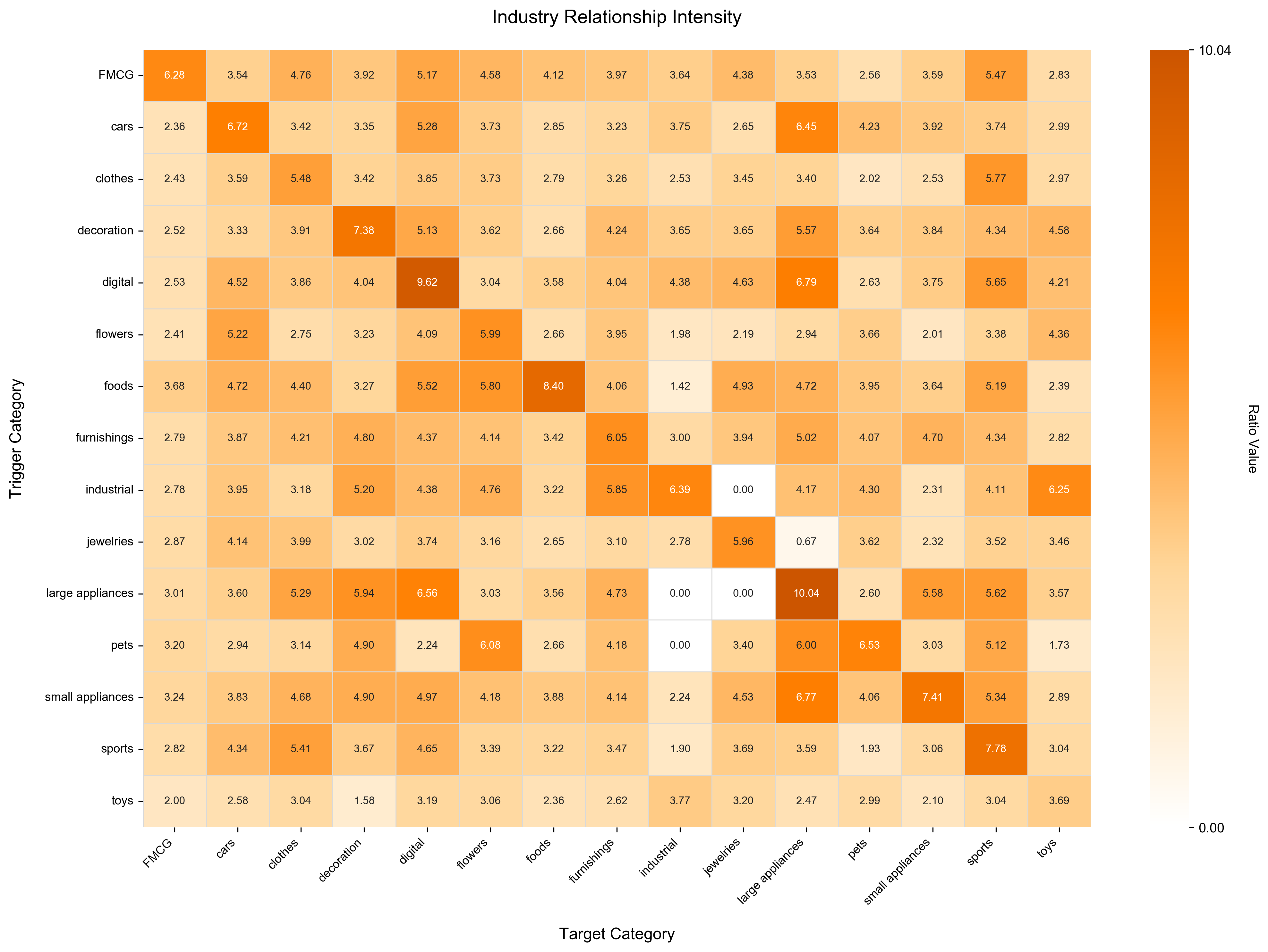}
\caption{The average click-through rate of different category pairs. }
\label{heatmap}
\end{figure}

As shown in Figure \ref{heatmap}, the average click-through rate of trigger and target items belonging to different categories is represented, where the horizontal and vertical axes represent the categories of target and trigger items, respectively. It can be seen that the CTR on the diagonal where trigger and target items share the same category is relatively higher than in other positions,
indicating that the relevance learning represents a critical signal in refining user interests and improving recommendation accuracy. To address these challenges, we propose a new method extracting the interaction and enhancing the relevance between trigger and target for the CTR prediction task in TIR scenarios. Specifically, we introduce the trigger-target interaction layer to explore cross relationships in an explicit-implicit hybrid manner of trigger and target items. Additionally, the cascading interest fusion module is proposed for adaptively fusing instant interest and personalized interest with the consideration of both predicted trigger intention and similarity. Furthermore, we introduce the pairwise loss guided with category-related information to explicitly enhance the learning of trigger relevance. Our contributions are summarized as follows:
\begin{itemize}
    \item We propose an explicit and implicit cross-gating layer and category-assisted pairwise loss to enhance the interaction and relevance between trigger and target.
    \item We propose the cascading interest fusion module to adaptively fuse instant interest and personalized interest with predicted trigger intention score and similarity in TIR scenarios. 
    \item The proposed CRRN method achieves the state-of-the-art performance on both industrial and public datasets, online A/B testing also demonstrates the effectiveness of our method.
\end{itemize}

\section{Method}

The architecture of CRRN is illustrated in Figure \ref{model}, which consists of three parts: 1) Trigger-Target Interaction layer is responsible for extracting explicit and implicit cross features between trigger and target, thus learning the complex relationship between them. 2) Cascading Interest Fusion module predictsthe user's trigger intention probability to fuse high-level features of trigger and target utilizing cascading attention blocks and cosine similarity. 3) Category-assisted Pairwise Loss incorporates category information of items from order relationships to emphasize trigger relevance at a more granular level. All the above features are concatenated and fed into Multi-Layer Perceptron (MLP) layers for predicting CTR probability.

CRRN takes five categories of features: 1) User Profile features including user id, user age, etc. 2) User Behaviors represents a list of items that users have clicked or purchased, reflecting historical preferences of the user; 3) Trigger features contain trigger id, trigger category, trigger brand and other attributes; 4) Target features contain the same information but of the target item; 5) Context features, such as timestamp and browsing position. Each feature is encoded as a one-hot vector and converted into a low-dimensional dense vector using embedding layer \cite{DIN}, and then subsequently concatenated to form the complete features. For convenience, the above features are respectively denoted as: $E_u$, $E_{seq}$, $E_t$, $E_i$ and $E_c$. Specially, $E_{seq}=\{E_{i}^{1},E_{i}^{2},...E_{i}^{L} \}$, where $L$ is the length of the behavior sequences.

\subsection{Trigger-Target Interaction}
In general, the fact that user clicks the trigger and enters the scrolling page, indicates a strong intent towards the trigger-related items. Therefore, exploring the complex relationships between trigger and target is essential for assessing whether the current item aligns with the user's immediate interests. Existing methods predominantly employ simple concatenation or direct multiplication operation to combine trigger and target features, which fail to capture the higher-level and complex cross relationships between trigger and target. To address this issue, the Trigger-Target Interaction layer first extracts the interaction features between the two items by element-wise multiplication, reinforcing their commonalities. Subsequently, based on the coaction features $E_{co}$, such as whether they belong to the same category, brand, or price level, a double-layer gating network is utilized to enhance and adaptively adjust the interaction features, ultimately yielding the refined interaction features $V_{in}$, which can be formulated as:
\begin{align}
& V_{in} = gate(E_{co}) \odot E_{t} \odot E_{i}, \\
& gate(x) = Sigmod(W_1\cdot Relu(W_0  x+b_0) + b_1) .
\end{align}
where $\odot$ denotes element-wise multiplication, and the project matrices $W_0$, $W_1$ are parameters to be learned.

\subsection{Cascading Interest Fusion}
In the TIR scenarios, extracting users' interests from their historical behavior sequences while simultaneously considering both trigger and target is crucial for CTR prediction tasks. Most existing methods use a simple concatenation of trigger and target attention features without considering their relationship, which often diminishes the role of triggers. Based on this consideration, we employ cascading attention blocks to adaptively fuse instant and personalized interests through predicted trigger intention scores and similarity, thus enhancing trigger relevance implicitly. 

\noindent
\textbf{Instant Intention Prediction.} As shown in the right part of Figure \ref{model}, user features, trigger features, and sum-pooled sequential features are fed into an MLP layer. Then the user's current trigger intention score can be obtained after sigmod activation function, which indicates the strength of user's interest in trigger-related items. The predicted trigger intent probability is formulated as:
\begin{align}
& \widehat{y}_{tri} = Sigmod(MLP([E_u \oplus E_t \oplus f_{sp}(E_{seq})]), \\
& \mathscr{L}_{tri} = - \frac{1}{N} \sum_{(x,y) \in S}^N (ylog(p(x))+(1-y)log(1-p(x)))  \label{loss equ}.
\end{align}
where $\widehat{y}_{int}$ is the predicted intention probability score,  $\oplus$ and $f_{sp}(\cdot)$ denotes the concatenation and sumpooling operation, respectively. $S$ represents the training set with size $N$, and $y$ is the trigger intention label. It is worth noting that the intention labels are page-level click indicators. Specifically, if the clicked item on the page shares the same category as the trigger, the intention label is marked as true, indicating a strong current intention towards the trigger.

\noindent
\textbf{Cascading Attention.} Then the cascading fusion module first performs personalized learning on the target item and historical behaviors using Multi-Head Attention mechanism \cite{trm} to extract multi-peaked user's historical behavior interests. Subsequently, it incorporates trigger information in a similar manner, thereby strengthening the importance of triggers and more comprehensively considering the triad of user, trigger, and target relationships. Take target attention for example:
\begin{align}
& V_{tar} = W^o[head_1^{tar} \oplus head_2^{tar} \oplus ...], \\
& head_i^{tar} = f_{attn}(W_q E_t, W_k E_{seq}, W_v E_{seq}) \\
& f_{attn}(Q, K, V) = softmax(\frac{Q  K}{\sqrt{d}}V) .
\end{align}
where $W_q, W_k, W_v$ are projection matrices of $head_i^{tar}$ to be learned. Trigger attention is processed in the same way but set the query and key as trigger features and target attention output respectively to obtain trigger attention output $V_{tri}$.

\noindent
\textbf{Interest Fusion.} Finally, based on aforementioned trigger intention score, instant and personalized interests are adaptively fused to obtain the final enhanced interest features $V_{fuse}$, the similarity between trigger and target is also utilized to revise the two parts considering the fact that trigger and target are increasingly similar, greater emphasis should be placed on the importance of trigger-relevant interest. $V_{fuse}$ is formulated as:
\begin{align}
V_{fuse}  = & \widehat{y}_{int} \cdot s(E_t,E_i) \cdot V_{tri} \\
& + (1-\widehat{y}_{int}) \cdot (1-s(E_t,E_i)) \cdot V_{tar}, \\
 s(E_t,E_i&)  = \frac{E_t^{T}E_i}{\Vert E_i^{T}\Vert \Vert E_i\Vert} .
\end{align}
where $s(\cdot,\cdot)$ is the cosine similarity to measure the similarity between trigger features and target features.

\subsection{Category-assisted Pairwise Loss}
Considering that attention-based modeling of trigger relevance is overly implicit, we introduce trigger relevance into the loss function to guide the process in estimating the click probability of candidate items. In contrast to traditional pairwise loss, which establishes a partial order relationship of clicks $>$ exposures, the category-assisted pairwise loss incorporates category relevance, establishing a new partial order relationship: clicks with relevance $>$ clicks without relevance $>$ exposures with relevance $>$ exposures without relevance. Here, relevance refers to target item sharing the same category with the trigger. 
\begin{equation}
\begin{split}
\mathscr{L}_{cpl} = & - \frac{1}{N^2} \sum_{i=1}^{N} \sum_{j=1}^{N} {(y_p(log(\widehat{y}_{ctr}^i}-\widehat{y}_{ctr}^j) \\
& + (1-y_p)log(1-(\widehat{y}_{ctr}^i-\widehat{y}_{ctr}^j))),
\end{split}
\end{equation}
where $y_p \in \{0,1\}$ is the order relationship label of item $i$ and item $j$. Finally, all features are fed into an MLP layer for the CTR prediction task, which is the negative log-likelihood function like equ \ref{loss equ}, and is denoted as $\mathscr{L}_{ctr}$. The total loss of CRRN is formulated as:
\begin{equation}
\mathscr{L} = \mathscr{L}_{ctr} + \lambda \mathscr{L}_{tri} + \theta \mathscr{L}_{cpl} . \label{loss equ} 
\end{equation}
where $\lambda$ and $\theta$ are hyper-parameters to balance the loss.

The CTR loss ensures the model accurately learn users' interests and preferences based on their historical behaviors, guaranteeing the overall correct learning direction of the training process. The intention loss learns the intensity of users' interest in the trigger, thereby precisely guiding and balancing instant and personalized interest in TIR scenarios. The category-assisted pairwise loss strengthens the trigger relevance explicitly by comparing category of item pairs. With the relevance and interaction enhancement between trigger and target items, our model can better capture users' interest changes in TIR scenarios, which boosts the performance of both the training and inferencing processes.

\section{Experiments}
\subsection{Experimental Setup}
\noindent
\textbf{Industrial Dataset:}
The industrial dataset is collected from the Tmall app, one of the largest shopping platforms, which focuses on branded products, capturing users' historical behaviors and log information. We selected samples from August 20, 2025, to October 20, 2025. This dataset includes about 500 million samples and each sample contains user, sequence, trigger, target, and contextual information. The validation and testing sets are sampled in October 20 and 21 respectively, containing about 7 million samples each. As for the label, the clicked item in the homepage recommendation is defined as the trigger item, which triggers a jump to the scrolling page named trigger-induced recommendation. The label is set to positive when the user click the item in the scrolling page, otherwise the label is set to 0.

\noindent\textbf{Public Dataset:} 
The public dataset\footnote{\url{https://tianchi.aliyun.com/dataset/dataDetail?dataId=56}} is sourced from Alibaba, collected by Alimama's advertising platform, involving 1 million users and 800 thousand ad items, totaling 26 million samples after processing. However, this dataset is not a strict TIR dataset, as it lacks trigger item information. 
Following the scheme in \cite{DIHN}, we define the latest item clicked within 4 hours before the sample as the trigger item, and samples that are not associated with a trigger will be discarded. 

\begin{table}[h!]
    \centering 
    \caption{Statistics of industrial and public datasets.} 
    \begin{tabular}{c ccc} 
        \hline 
        Dataset & Users & Items & Samples \\
        \hline
        Industrial Dataset  &  7,692,277  & 22,512,661  & 449,786,531 \\
        Public Dataset  &  1,366,056  & 846,812  & 26,557,962 \\
        \hline
    \end{tabular} 
    \label{ablaTable}
\end{table}

\noindent\textbf{Implementation details:}
We use Adagrad \cite{adagrad} optimizer with decay learning rate, which is initialized as $0.001$,and the decay rate is set as $0.95$. Our batch size is $1024$, and the training epoch is set to $4$, $\lambda$ and $\theta$  in equ \ref{loss equ} are experimentally set as $0.2$ and $0.5$, respectively to balance the loss. The sequence length is set as 100 to cover the majority sequence, where insufficient parts are masked, and excess parts are trimmed. As Area Under Curve (AUC) \cite{DIN} and RelaImpr are widely used for the CTR prediction task, we adopt them as the main metrics for the experimental evaluation. 
As for the intention label, it serves as a supervisory signal that is exclusively utilized during training and is not employed while inferencing.

\subsection{Performance Comparison}
We compare the proposed CRRN with current state-of-the-art methods: Wide\&Deep+TIR \cite{wide&deep}, DIN+TAR\cite{DIN}, DIHN \cite{DIHN}, DIAN \cite{DIAN} and DEI2N \cite{DEI2N}. For fair comparison, we equip Wide\&Deep+TR and DIN with instant interest modeling using trigger-related features. 
\begin{itemize}
    \item \textbf{Wide\&Deep+TIR} jointly trained wide linear models and deep neural networks for memorization and generalization in recommender systems. Trigger features are added as input to capture instant interest.
    \item \textbf{DIN+TAR} utilizes attention mechanism to adaptively learn the representation of user interests from historical behaviors with both trigger and target.
    \item \textbf{DIHN} is the first to introduce the TIR task and employs the deep neural network to learn user's intention for better recommendation in TIR scenarios.
    \item \textbf{DIAN} proposes an intention-aware network that identifies user's intention by dynamically balancing the results of trigger-free and trigger-aware recommendations.
    \item \textbf{DEI2N} introduces an instant interest layer aimmimg at learning the dynamic changes in the intensity of instant interest after users click on the trigger item, therefore enabling the modeling of finer-grained interest.
\end{itemize}

\begin{table}[h!]
    \centering 
    \caption{Comparison AUC results on industrial and public datasets.} 
    \vspace{0.5em}
    \begin{tabular}{c cc cc} 
        \hline 
        \multirow{2}{*}{Model}   & \multicolumn{2}{c}{Industrial} & \multicolumn{2}{c}{Public}  \\
        \cline{2-5}
        & AUC & RelaImpr & AUC & RelaImpr \\
        \cline{1-5}
        Wide\&Deep+TR  & 0.6634  & -0.85\%  &  0.6310 & -3.03\% \\
        DIN+TRA      & 0.6648  & 0.00\%   &  0.6351 & 0.00\% \\
        \hline
        DIHN         & 0.6664  & +0.97\%  &	0.6388  & +2.74\% \\
        DIAN         & 0.6693  & +2.73\%  & 0.6408  & +4.22\% \\
        DIE2N        & 0.6723  & +4.55\%  &	0.6428  & +5.70\% \\
        \hline    
        CRRN         & 	\textbf{0.6752} & \textbf{+6.31\%}  &		\textbf{0.6448}  & \textbf{+7.18\% } \\
        \hline
    \end{tabular} 
    \label{compTable}
\end{table}

The comparison results are presented in Table \ref{compTable}. It can be seen that trigger-based series methods reveal significant improvements over traditional methods like Wide\&Deep and DIN, highlighting the importance of trigger information and intention estimating in TIR scenarios. DEI2N achieves the best results besides ours, indicating the necessity of extracting interaction features. Notably, CRRN outperforms all other methods, as it comprehensively utilizes the complex relationships between trigger and target items, and also enhances relevance representation through cascading attention blocks and category-assisted order relationship learning, enabling more accurate balancing of instant and personalized interests.

\subsection{Ablation Study}
We also conduct several ablation experiments to evaluate the effectiveness of each module in CRRN, as shown in Table \ref{ablaTable}, where the baseline is DIN+TRA, serving as the previous online model. We observe that each module contributes to improving the model's prediction accuracy to some extent. The Trigger-Target Interaction (TTI) layer enhances CTR estimation by $0.54\%$ from learning the complex explicit and implicit cross relationships between trigger and target, providing more comprehensive features for the final task. The Cascading Interest Fusion (CIF) module improves performance by $0.78\%$, which assists the model in determining and balancing instant and personalized interests. Category-assisted Pairwise Loss (CPL) addresses the insufficient exploring on trigger relevance and refines the ability to learn category-specific relationships between each item pairs, bringing a $0.43\%$ increase. All the modules contribute to better CTR predictions.
\begin{table}[h!]
    \centering 
    \caption{Ablation AUC results on industrial dataset.} 
    \begin{tabular}{c ccc c c} 
        \hline 
        Model & TTI & IEF & CPL & AUC & RelaImpr \\
        \hline
        Baseline      &  \XSolidBrush  & \XSolidBrush   & \XSolidBrush   & 0.6648  & 0.00\% \\
        B+TTI      & \Checkmark & \XSolidBrush   & \XSolidBrush   & 0.6702 & 3.27\% \\
        B+CIF      & \XSolidBrush & \Checkmark & \XSolidBrush   & 0.6726  & 4.73\% \\
        B+CPL       & \XSolidBrush & \XSolidBrush  & \Checkmark & 0.6691 & 2.61\%  \\
        \hline    
        CRRN          & \Checkmark & \Checkmark & \Checkmark & \textbf{0.6752} &  \textbf{+6.31\%}  \\
        \hline
    \end{tabular} 
    \label{ablaTable}
\end{table}

\subsubsection{Ablation of Intention Label}
In recommendation systems, category maintains a balanced relevance without being overly focused, aligning with the similarity requirements of the TIR scenario and intuitively reflecting users' intention to some extent. Additionally, we conduct an ablation study by treating brands, shops and price-level as labels in the CPL module. From Table \ref{ablaTabl-label}, it is evident that category exhibits the best performance, with others demonstrating relatively limited impact. This finding is consistent with the feature importance analysis, which indicates that category coaction gets the highest important score among category/price-level/brand/shop coaction features (0.0071, 0.0026, 0.0011, 0.00079, respectively). 
This may be attributed to the fact that products from the same brand or shop tend to be more diversified, failing to adequately meet the requirements of instant interest modeling.
\begin{table}[h!]
    \centering 
    \caption{Ablation AUC results of intention label.} 
    \begin{tabular}{c c c} 
        \hline 
        Model &  AUC & RelaImpr \\
        \hline
        Baseline      & 0.6648  & 0.00\% \\
        CRRN-price      & 0.6735  & 5.28\% \\
        CRRN-brand    & 0.6721 & 4.43\% \\
        CRRN-shop      & 0.6718  & 4.25\% \\
        CRRN-category      & \textbf{0.6752} &  \textbf{+6.31\%}  \\
        \hline
    \end{tabular} 
    \label{ablaTabl-label}
\end{table}

\subsubsection{Ablation of Similarity Measurements}
Cosine similarity is commonly used in recommendation systems to measure the similarity between two vectors in high-dimensional space. We also explore different methods to assess the similarity between trigger and target items as the strength of trigger intention, including cosine similarity, MLP, bilinear, and contrastive learning pretraining. From Table \ref{ablaTabl-sim}, we can observe that cosine similarity yields the best performance, likely because it can effectively preserve the original characteristics of items to a greater extent, which is relatively important for users' decision-making. On the other hand, learnable parameters in other methods may capture high-dimensional features, potentially deviating from the fundamental relevance.
\begin{table}[h!]
    \centering 
    \caption{Ablation AUC results of similarity measurements.} 
    \begin{tabular}{c ccc c c} 
        \hline 
        Model & AUC & RelaImpr \\
        \hline
        Baseline      & 0.6648  & 0.00\% \\
        CRRN-MLP        & 0.6739 & 5.52\% \\
        CRRN-bilinear      & 0.6747  & 6.01\% \\
        CRRN-CL      & 0.6743  & 5.76\% \\
        CRRN-cosine         & \textbf{0.6752} &  \textbf{+6.31\%}  \\
        \hline
    \end{tabular} 
    \label{ablaTabl-sim}
\end{table}

\subsection{Time Analysis}
We compare the training and inference efficiency of CRRN with other methods. The epoch number is set to 1. The results in Table \ref{time} indicate that CRRN's efficiency is comparable to these baselines. Wide\&Deep+TR, which does not employ attention sequence modeling, has the lowest time consumption. Since DIHN, DIAN, DEI2N, and CRRN all incorporate additional modules to explore trigger-related interests, their time consumption is slightly higher than DIN+TRA by approximately 7-10 minutes. Consequently, the additional MLP for learning trigger interaction features and the varying degrees of trigger intention do not introduce much more time cost. In terms of online inference latency, CRRN is deployed on an NVIDIA L20 GPU, where it only adds 2ms of time consumption compared to the baseline, fully satisfying our online service requirements of 150 milliseconds.

\begin{table}[h!]
    \centering 
    \caption{Execution time (minutes) Results.} 
    \begin{tabular}{c ccc} 
        \hline 
        Method & Training & Prediction & Online \\
        \hline
        Wide\&Deep+TR  & 102  & 8  & - \\
        DIN+TRA      & 127  & 10   & 42ms  \\
        DIHN         & 134  & 11   & - \\
        DIAN         & 137  & 11   & - \\
        DIE2N        & 135  & 11   & - \\
        \hline    
        CRRN         & 	134 & 11 & 45ms \\
        \hline
    \end{tabular} 
    \label{time}
\end{table}

\subsection{Online A/B Testing}
To further validate the effectiveness of CRRN method, we also conduct online A/B Testing from April 10th to 17th on the platform of Tmall. During this evaluation period, we compare CRRN against DIN with trigger attention, which serves as the previous online model. The results are represented in Table \ref{online}. 
PV and IPV are short for Page View, Item Page View, respectively, where PV and IPV are the average numbers of items that users browse and click in the TIR scenario. 
We can observe that CRRN contributes up to $3.87\%$ pCTR promotion, which denotes the average number of items that the user clicks on the TIR scenario, demonstrating the effectiveness of our proposed method. And now CRRN is serving online.

\begin{table}[h!]
    \centering 
    \caption{Online A/B testing results.} 
    \begin{tabular}{c ccc} 
        \hline 
        Method & IPV & PV & CTR \\
        \hline
        DIN+TAR  &  0.87  & 19.70  & 3.11\% \\
        CRRN  &  0.92  & 20.34  & 3.23\% \\
        \hline
        Lift rate & +5.75\% & +3.25\% & +3.87\% \\
        \hline
    \end{tabular} 
    \label{online}
\end{table}

\subsection{Case Study}
\begin{figure}[htpb]
\centering
\includegraphics[scale=0.30]{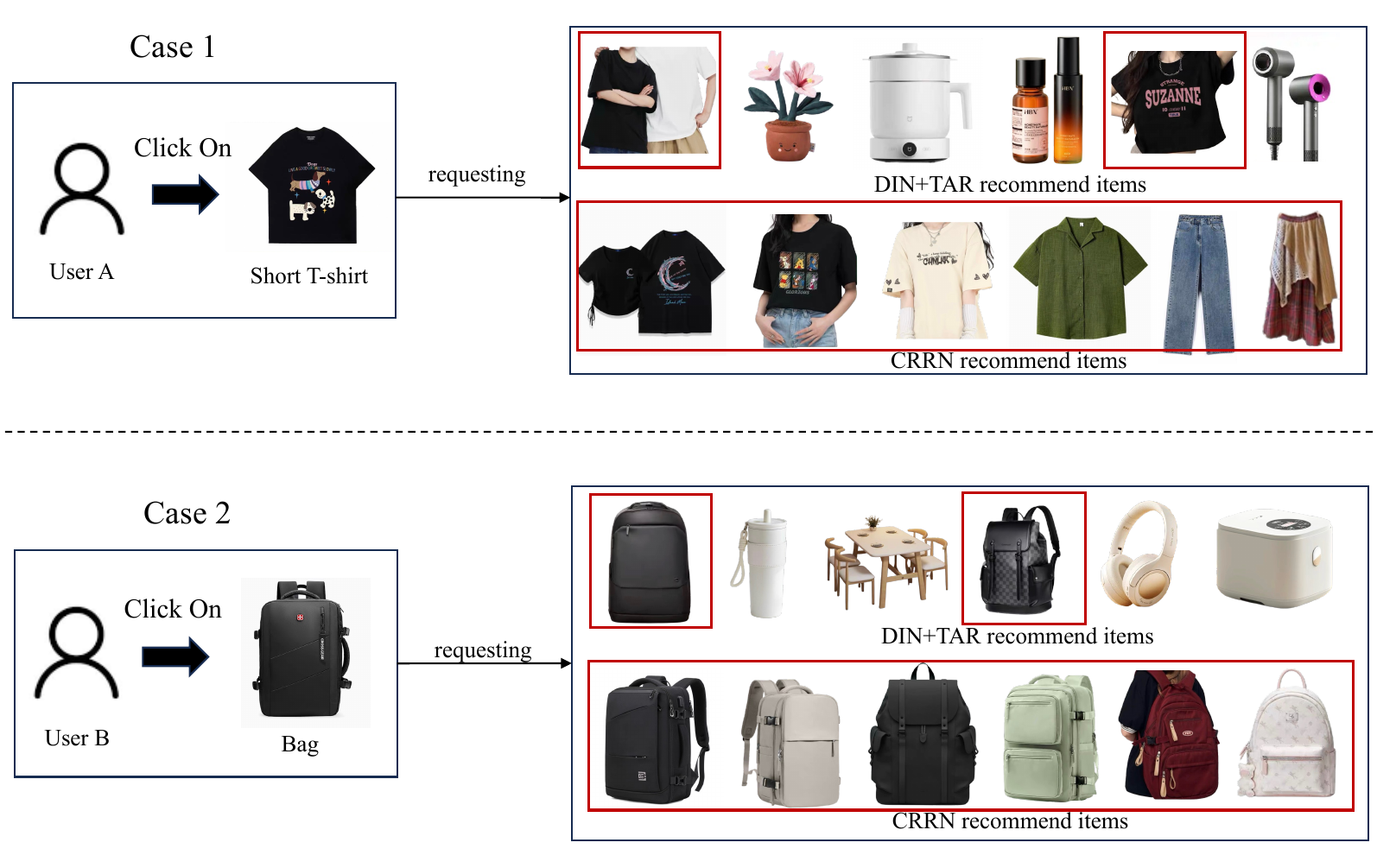}
\caption{Two cases of CRRN and compared baseline.}
\label{case}
\end{figure}

We demonstrate the enhancement of relevance through two real-world recommendation examples. As shown in Figure \ref{case}, the upper part represents a user clicking on a short T-shirt that is defined as the trigger item in the TIR scenario. With DIN+TAR, the recommended item categories are relatively scattered, primarily consisting of items that the user has recently interacted with (clicks or adds to cart), showing weaker relevance to the trigger item. This limits the system's ability to effectively capture and reinforce the user's immediate interest, thereby impacting the immersive shopping experience. In contrast, CRRN predominantly recommends trigger-related items, including various short T-shirts and complementary clothing such as pants and dresses, which provide users with a more immersive and focused shopping experience. Another example further supports this observation.

\subsection{Discussion}
We conduct a more fine-grained categorization of categories and statistically analyze the recommendation results of CRRN online to explore what makes CRRN better. The data in the table \ref{discussion} shows that compared to the base model, the proportion of recommended items in the Same Category (SC) and Related Category (RC) has increased, which can be attributed to the importance of relevance in instant interest modeling within the TIR scenario. CRRN addresses this by learning the complex relationship between trigger and target items through implicit feature interaction and explicit relevance enhancement. Moreover, table \ref{discussion} also reveals a certain decrease in the proportion of items from Different Category (DC), though they still maintain a significant share of $11.23\%$. This is because the balance between relevance and personalization is crucial for users' experience. A weak relevance may lead to overly scattered recommendations, failing to meet users' instant interests, while a strong relevance can result in a screen filled with similar items, causing user fatigue. CRRN addresses this challenge by leveraging supervised intention modeling to strike a balance between instant interests and personalized interests, effectively capturing users' intention in TIR scenarios while satisfying certain personalized demands.

\begin{table}[h!]
    \centering 
    \caption{Fine-grained category relevance statistics online.} 
    \begin{tabular}{c ccc} 
        \hline 
        Method & SC & RC & DC \\
        \hline
        DIN+TAR  &  40.51\%  & 7.16\%  & 52.33\% \\
        CRRN  &  67.45\%  & 21.31\%  & 11.24\% \\ 
        \hline
        diff & +26.94\% & +14.15\% & -41.09\% \\
        \hline
    \end{tabular} 
    \label{discussion}
\end{table}

As for generalization, CRRN is applicable to TIR scenarios and related or similar recommendations, which are widely adopted across various domains such as e-commerce, news, and video software. For example, Taobao, Amazon, and related recommendations in news, video apps, etc.

\section{Conclusion}
In this paper, we propose an effective method CRRN for the CTR prediction task in TIR scenarios, emphasizing the importance of interaction and relevance between trigger and target. The method consists of three key components: the Trigger-Target Interaction layer learns both explicit and implicit cross relationships between trigger and target items through personalized gate, thus providing more comprehensive features for subsequent decision-making. The Cascading Interest Fusion module predicts the user's trigger intention and adaptively fuses instant and personalized interests. The Category-assisted Pairwise Loss further enhances trigger relevance learning with the guidance of category association, thereby better adapting to users' immersive experience needs in TIR scenarios. Extensive experiments on both industrial and public datasets, as well as online A/B testing demonstrate the effectiveness of our approach.

\clearpage
\bibliographystyle{IEEEtranS}
\bibliography{ref}

\end{document}